\documentclass[reprint, amsmath,amssymb,aps,longbibliography]{revtex4-1}

 \usepackage[bottom]{footmisc}

\usepackage{kpfonts}
\usepackage[T1]{fontenc}
\usepackage[latin9]{inputenc}
\usepackage{mathrsfs}
\usepackage{bm}
\usepackage{amsmath}
\usepackage{amsthm}
\usepackage{amssymb}
\usepackage{stmaryrd}
\usepackage{mathtools}

\usepackage{stackengine}
\stackMath
\newcommand{\widesim}[2][1.5]{
  \mathrel{\overset{#2}{\scalebox{#1}[1]{$\sim$}}}
}

\edef\ordinarycolon{\mathchar\the\mathcode`: }
\edef\ordinaryequals{\mathchar\the\mathcode`= }

\usepackage{breqn}
\catcode`^=7

\AtBeginDocument{%
  \catcode`^=12
}

\makeatletter

\allowdisplaybreaks

\let\cat@comma@active\@empty

\usepackage[capitalize]{cleveref}

\usepackage{color}
\RequirePackage[dvipsnames,usenames]{xcolor}

\newif\ifnotes
\notestrue

\renewcommand{\mid}{\vert}

\makeatother

\newcommand{\ba}{\begin{eqnarray}}
\newcommand{\ea}{\end{eqnarray}}

\newcommand{\eq}[1]{\begin{align}#1\end{align}}

\newcommand{\II}{{\mathcal{I}}}

\newcommand{\AAA}{{\mathcal{A}}}
\newcommand{\BBB}{{\mathcal{B}}}

\newcommand{\R}{{\mathbb{R}}}

\newcommand{\NN}{\mathcal{N}}

\newcommand{\vx}{{\pmb{x}}}
\newcommand{\vX}{{\pmb{X}}}
\newcommand{\oo}{{\omega}}
\newcommand{\vP}{{\bold{P}}}

\begin{document}

\title{Fluctuation theorems for multiple co-evolving systems}

\author{David H. Wolpert}
\affiliation{Santa Fe Institute, Santa Fe, New Mexico \\
Complexity Science Hub, Vienna\\
Arizona State University, Tempe, Arizona\\
\texttt{http://davidwolpert.weebly.com}}

\begin{abstract}
All previously derived thermodynamic fluctuation theorems (FTs) that concern multiple co-evolving systems have
required that each system can only change its state during an associated pre-fixed, limited set of times.
However, in many real-world cases the times when systems change their states are randomly determined, e.g., in almost all 
biological examples of co-evolving systems. Such randomness in the timing
drastically modifies the thermodynamics.
Here I derive FTs that apply whether or not the timing is random.
These FTs provide new versions of the second law, and of \textit{all} conventional thermodynamic uncertainty relations (TURs). 
These new results are often stronger than the conventional
versions, which ignore how an overall system may decompose into a set of co-evolving systems.
In addition, the new TURs often bound entropy production (EP) of the overall system even if
none of the criteria for a conventional TUR (e.g., being a nonequilibrium steady state) hold for that overall system.
In some cases the new FTs also provide a new type of speed limit theorem, and in some cases they also
provide nontrivial \textit{upper} bounds on expected EP of systems.
I use a standard example of ligand detectors in a cell wall to illustrate these results.

\end{abstract}

\maketitle

\section{Introduction}
Some of the most important results in stochastic
thermodynamics are the fluctuation theorems (FTs)~\cite{seifert2012stochastic,van2015ensemble,Bisker_2017,esposito2010three,wolpert_thermo_comp_review_2019}. These
govern the probability distribution of the total entropy production (EP) generated over any 
time interval by a system that evolves according to a continuous-time Markov chain (CTMC). 
As an example of their power, the FTs provide bounds strictly more powerful than 
the second law, leading them to be identified as the ``underlying
thermodynamics ... of time's arrow''~\cite{seif2020machine}. In addition,
an FT can be used to derive a thermodynamic uncertainty relation (TUR)~\cite{hasegawa2019fluctuation}, i.e.,  
an upper bound on how precise \textit{any} current can be
in an evolving system in terms of the expected EP it generates.

Many evolving systems decompose into a set of multiple co-evolving systems. Examples include
a digital circuit, which decomposes into a set of interacting 
gates~\cite{wolpert_thermo_comp_review_2019,wolpert2018thermo_circuits}, and 
a cell, which decomposes into a set of  
many organelles and biomolecule species, jointly
evolving as a multipartite process (MPP)~\cite{horowitz2014thermodynamics,horowitz_multipartite_2015}.
The early work on FTs did not consider the thermodynamic implications of such a decomposition.
While some recent papers have considered those 
implications ~\cite{hartich_sensory_2016,horowitz2014thermodynamics,van2020thermodynamic,wolpert2020minimal,ito2020unified},
they have not derived FTs.
(One exception is~\cite{hartich_stochastic_2014}, which does derive an FT --- but that FT only applies to bipartite systems,
and only if the system is in a nonequilibrium stationary state (NESS) \footnote{\cite{shiraishi_ito_sagawa_thermo_of_time_separation.2015,shiraishi2015fluctuation}
provide an FT for a bipartite system which holds even if the system isn't in an NESS. However,
this FT does \textit{not} involve the conventional definition of $\sigma_i$, the ``entropy production'' of a subsystem $i$, as the net 
entropy change of the entire universe over all instances when subsystem $i$ changes its state. Due to this,
their $\sigma_i$ can have negative expected values,
and we cannot use the associated FT to derive TURs, as done for example using the FT for 
conventionally defined EP in~\cite{hasegawa2019fluctuation}. 
Similarly, we cannot use their FT involving that $\sigma_i$ to derive conditional FTs, like the conditional FTs introduced below.
See App.\,C in~\cite{wolpert2020uncertainty}.
}.)

There has also been a line of papers that derive FTs for 
an arbitrary number of co-evolving systems, which apply without any NESS 
restriction~\cite{ito2013information,ito_information_2015,wolpert.thermo.bayes.nets.2020}. However, these papers all require
that the timing sequence $\mathcal{T}$ specifying when each system changes its state is 
fixed ahead of time, not randomly determined as the full system evolves. This requirement implicitly assumes a global clock,
which simultaneously governs all the systems, synchronizing their dynamics \footnote{There
are also several analyses in which $\mathcal{T}$ is fixed, and in addition the dynamics is a discrete-time Markov
chain, with all systems updating simultaneously, e.g.,~\cite{bo2015thermodynamic}. In general, such
discrete-time chains cannot even be approximately implemented as a CTMC~\cite{owen_number_2018,wolpert_spacetime_2019}. 
This results in a more fundamental distinction between such models and those in which $\mathcal{T}$ is random.
}.

In many real-world scenarios however --- arguably in almost all biological scenarios ---
the timings $\mathcal{T}$ are random. 
Moreover, the thermodynamics changes drastically depending on whether
$\mathcal{T}$ is random or fixed. 
For example, in a fixed-$\mathcal{T}$ scenario,
there is a pre-determined time-interval assigned to each system, specifying when its state can change.
The global rate matrix must change from any one such interval to the next.
So thermodynamic work must be done on any such fixed-$\mathcal{T}$ system, simply to enforce the sequence
$\mathcal{T}$. No such work is required when $\mathcal{T}$ is random --- such systems can have
time-homogeneous rate matrices.

Here I fill in this gap in stochastic thermodynamics, by deriving FTs that hold for arbitrary MPPs,
whether or not $\mathcal{T}$ is random. 
These FTs provide new versions of the second law and of (all) conventional TURs
(including, e.g.,~\cite{horowitz_gingrich_nature_TURs_2019,liu2020thermodynamic,hasegawa2019fluctuation,falasco2020unifying,barato2015thermodynamic}), which
are often stronger than those earlier, conventional TURs. 
In particular, I derive TURs that bound EP of the full system in terms of current precision(s) of its 
constituent co-evolving systems. These
TURs apply even if the full system does not meet the criteria for a conventional TUR 
(e.g., being an NESS), so long as at least one of the constituent systems meets one of those criteria. 
Moreover, in some cases different TURs apply to different  constituent systems,
and can be combined to lower-bound the EP of the full system in terms of current precisions of those systems. 
In addition, in some cases the new FTs provide a new type of speed limit theorem~\cite{zhang2018comment,shiraishi_speed_2018,okuyama2018quantum,van2020unified}, involving
changes in conditional mutual information from the beginning to the end of a process rather than the distance
between initial and final distributions. Finally, the new FTs also sometimes provide 
information-theoretic \textit{upper} bounds
on expected EP of systems.
%
I end by illustrating these results
in
a standard example of receptors in a cell wall sensing their external environment.
All proofs and formal definitions not in the text are in the Supplemental Material at
[URL will be inserted by publisher].

\section{Multipartite Stochastic Thermodynamics Elementary Concepts}
\label{sec:terminology}

$\NN$ is a set of $N$ systems, with finite state spaces
$\{X_i : i = 1, \ldots N\}$. $x$ is a vector in $X$, the joint space of $\NN$, and
$\vX$ is the set of all trajectories $\vx$ of the joint system.
As in~\cite{horowitz_multipartite_2015,wolpert2020minimal}, I assume that each system is 
in contact with its own reservoir(s), and
so the probability is zero that any two systems change state simultaneously.
Therefore there is a set of time-varying stochastic rate matrices, 
$\{K^{x'}_x(i; t) : i = 1, \ldots, N\}$, where for all $i$, $K^{x'}_x(i; t) = 0$ if $x'_{-i} \ne x_{-i}$, and
the joint dynamics is given by~\cite{horowitz2014thermodynamics,horowitz_multipartite_2015}
\eq{
\frac{d p_x(t)}{dt} &= \sum_{x'} K^{x'}_{x}(t) p_{x'}(t)   =  \sum_{x'} \sum_{i \in \NN} K^{x'}_{x}(i; t) p_{x'}(t)
}
For any $A \subseteq \NN$ I define $
K^{x'}_{x}(A; t) := \sum_{i \in A} K^{x'}_{x}(i; t).$

\begin{figure}[tbp]
\includegraphics[width=75mm]{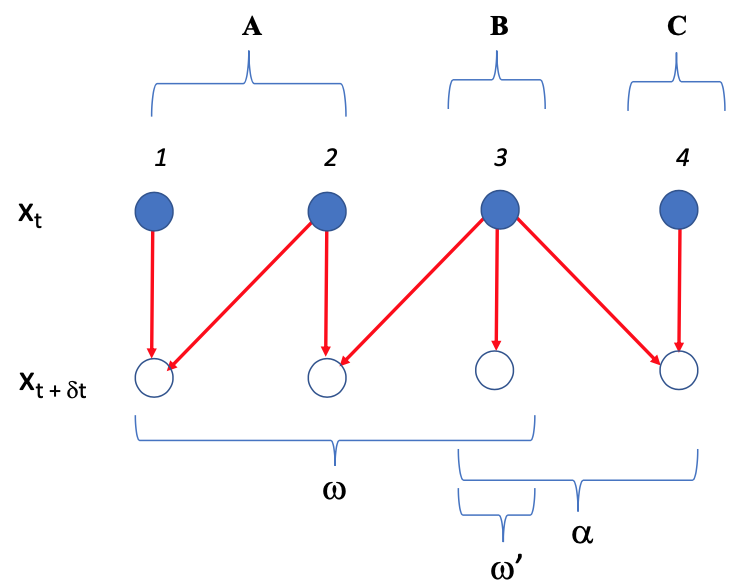}
\caption{Four systems, $\{1, 2, 3, 4\}$ interacting in a MPP.
The red arrows indicate dependencies in the associated four rate matrices. 
$B$ evolves independently, but is continually observed by $A$ and $C$. So the statistical coupling
between $A$ and $C$ could grow with time, even though their rate matrices do not involve one
another. Three examples of units are the sets $\oo, \oo', \alpha$
indicated at the bottom: $r(1) = r(2) = \oo$,
$r(4) = \alpha$, and $r(3) = \oo'$. 
}
\label{fig:1}
\end{figure}

For each system $i$, $r(i; t)$ is any set of systems at time $t$ that includes $i$ such that
we can write
\eq{
K^{x'}_x(i; t) = K^{x'_{r(i;t)}}_{x_{r(i;t)}}(i; t) \delta(x'_{-r(i;t)}, x_{-r(i;t)})
\label{eq:def_unit_rate}
} 
for an appropriate set of functions $K^{x'_{r(i;t)}}_{x_{r(i;t)}}(i; t)$.  In general, for any given $i$, there
are multiple such sets $r(i;t)$.
A \textbf{unit} $\oo$ (at an implicit time $t$) is a set of systems such that $i \in \oo$ implies that $r(i;t) \subseteq \oo$.
Any intersection of two units is a unit, as is any union of two units. 
(See  \cref{fig:1} for an example.)
A set of units that covers $\NN$ and is closed under intersections is a \textbf{(unit) dependency structure}, typically written as $\NN^*$.
Except where stated otherwise, I focus on dependency structures which do not include $\NN$ itself as a member. 
From now on I assume there are pre-fixed time-intervals in which $\NN^*$ doesn't change,
and restrict attention to such an interval. (This assumption holds in all of the papers mentioned above.)

For any unit $\oo$, I write $
K^{x'_\oo}_{x_\oo}(\oo; t) := \sum_{i \in \oo} K^{x'_{\oo}}_{x_{\oo}}(i; t)$.
So $K^{x'}_x(\oo; t) := \sum_{i \in \oo} K^{x'}_{x}(i; t) = K^{x'_\oo}_{x_\oo}(\oo; t)  \delta(x'_{-\oo}, x_{-\oo})$.
Crucially, at any time $t$, for any unit $\omega$, $p_{x_{\omega}}(t)$ evolves as a self-contained CTMC with rate
matrix $K^{x'_{\omega}}_{x_\omega}(\omega; t)$:
\eq{
\frac{d p_{x_\oo}(t)}{dt} 
	&= \sum_{x'_\oo} K^{x'_\oo}_{x_\oo}(\oo; t) p_{x'_\oo}(t)
\label{eq:15aa}
}
(See App.\,A in~\cite{wolpert2020minimal} for proof.) Therefore any unit obeys all the usual stochastic thermodynamics
theorems, e.g., the second law, the FTs, the TURs, etc. In general
this is not true for a single system in an MPP~\cite{wolpert2020minimal}.

As an example,~\cite{hartich_sensory_2016,bo2015thermodynamic} considers an MPP where receptors in a cell wall observe the
concentration of a ligand in a surrounding medium, without any back-action from the receptor onto that concentration level~\cite{sagawa2012fluctuation,van2020thermodynamic,horowitz2011designing,shiraishi_ito_sagawa_thermo_of_time_separation.2015,wachtler2016stochastic,verley_work_2014}. In addition to the medium and 
receptors, there is a memory that observes the state of those receptors. We can extend that scenario, to include a second set of
receptors that observe the same medium (with all variables appropriately coarse-grained).  
\cref{fig:1} illustrates this MPP; system $3$ is the concentration level,
system $2$ is the first set of receptors observing that concentration level, system $1$ is the memory, 
and system $4$ is the second set of receptors.



Taking $k_B = 1$, I write the inverse temperature of reservoir $k$ for system $i$ as $\beta_i^k$,
 the associated chemical potentials as $\mu_i^k$ (with  $\mu_i^k = 0$ if $k$ is a heat bath), and
any associated conserved quantities as $n_i^k(x_i)$.
So the rate matrix of system $i$ is $
K_x^{x'}(i; t) = \sum_k K_x^{x'}(i; k, t)$.
%
Any fluctuations of $x$ in which only $x_i$ changes  are determined by exchanges between $i$ and its reservoirs.
Moreover, since we have a MPP, the rate matrices $K(j;t)$ for $j \ne i$ 
must equal zero for such a fluctuation in the state of $i$. Therefore, writing  $H_{x}(t)$ for the
\textbf{global} Hamiltonian, thermodynamic consistency~\cite{rao2018conservation,van2013stochastic} says that for all $i, k, t, x, x'$,
\eq{
\!\!\!\! \!\!\!\! \!\!\!\! \ln \dfrac{K^{x'}_{x}(i; k, t)}{K^{x}_{x'}(i; k, t)} 
	= \beta_i^k\left(\left[H_{x'}(t) - H_{x}(t)\right] + \mu_{i}^k\left[n_i^k(x_i) - n_i^k(x'_i)\right]\right)
\label{eq:nldb1}
}
so long as both $x_{-i} = x'_{-i}$ and $K^{x'}_{x}(i; k, t) \ne \delta(x'_i, x_i)$.

The LHS of \cref{eq:nldb1} cannot depend on values $x_j$ or $x'_j$ for $j \not \in r(i)$. Since we are only
ever interested in differences in energy, this means that we must be able to rewrite
the RHS of \cref{eq:nldb1} as
\eq{
\beta_i^k\left(\left[H_{x'_{r(i)}}(i;t) - H_{x_{r(i)}}(i;t)\right] + \mu_{i}^k\left[n_i^k(x_i) - n_i^k(x'_i)\right]\right)
\label{eq:nldb}
}
for some \textbf{local} Hamiltonians $H_{x_{r(i)}}(i; t)$.  (Some general thermodynamic consequences of this 
 \textbf{system} LDB (SLDB) condition are discussed in the Supplemental Material at
[URL will be inserted by publisher].)

To define trajectory-level quantities, first, for any set of systems $\alpha$,
define the \textbf{local} stochastic entropy 
as
\eq{
s^{\alpha}(\vx_\alpha(t)) &:= -\ln p_{\vx_\alpha(t)}(t) 
\label{eq:stoch_entropy}
}
In general I will use the prefix $\Delta$ to
indicate the change of a variable's value from $t = 0$ to $t = t_f$, e.g., 
\eq{
\Delta s^{\alpha}(\vx_\alpha) &:= \ln p_{\vx_\alpha(0)}(0) - \ln p_{\vx_\alpha(t_f)}(t_f)
\label{eq:43}
}
The trajectory-level entropy flows (EFs) between a system $i$ and its reservoirs
along a trajectory is written as $Q^i(\vx)$.
The \textbf{local} EF into $\alpha$ is the total EF into the systems $\alpha$ from their reservoirs during $[0,t_f]$:
\eq{
Q^\alpha(\vx) &:= \sum_{i \in \alpha} Q^i(\vx)
\label{eq:19}
}

The \textbf{local} EP of any set of systems $\alpha$ is 
\eq{
\sigma^\alpha(\vx) &:=  \Delta s^\alpha(\vx) - Q^\alpha(\vx)
\label{eq:local_EP}
}
which can be evaluated by combining \cref{eq:43} and the expansion of local EF in the Supplemental Material at
[URL will be inserted by publisher].
For any unit $\oo$, the expectation of $\sigma^\oo$ is
non-negative. (This is not true for some quantities called ``entropy production'' in the literature,
nor for the analogous expression ``$\sigma_x$'' defined just before Eq.\,(21) in~\cite{shiraishi2015fluctuation}.)
Due to the definition of a unit, the definition of $Q^i(\vx)$, and \cref{eq:19},
the EF into (the systems in) $\oo$ along trajectory $\vx$ is only a function of $\vx_\oo$.
So we can write $Q^\oo(\vx) = Q^\oo(\vx_\oo)$. 
Since $\Delta s^\oo(\vx)$ also only depends on $\vx_\oo$, 
we can write $\sigma^\oo(\vx)$ as $\sigma^\oo(\vx_\oo)$.

Setting $\alpha = \NN$ in \cref{eq:43,eq:local_EP} allows us to
define {global} versions of those trajectory-level thermodynamic quantities. In particular, the \textbf{global} EP is
\eq{
\sigma^\NN(\vx) &:=  \Delta s^\NN(\vx) - Q^\NN(\vx)
\label{eq:global_EP}
}
There are several ways to expand the RHS of \cref{eq:global_EP}.
One of them, discussed in the Supplemental Material at
[URL will be inserted by publisher],
involves an extension of what is called the ``learning rate''
of a bipartite process~\cite{barato_efficiency_2014,hartich_stochastic_2014,hartich_sensory_2016,matsumoto2018role,Brittain_2017}, to 
MPPs with more than two systems~\cite{wolpert2020minimal}.
Here  
I focus on a
different decomposition.

To begin, let ${\NN^*} = \{\oo_j : j = 1, 2, \ldots, n\}$ be a dependency structure.
Suppose we have a set of functions indexed by the units, $\{f^{\oo_j} : \vX \rightarrow \R\}$. The
associated \textbf{inclusion-exclusion sum} (or just ``in-ex sum'') is 
defined as
\eq{
\widehat{\sum_{\oo' \in {\NN^*}}} f^{\oo'}(\vx) &:= \sum_{j = 1}^n f^{\oo_j}(\vx) - \sum_{1 \le j < j' \le n} f^{\oo_j \cap \oo_{j'}}(\vx) \nonumber \\
	& \qquad+ 	\sum_{1 \le j < j' < j'' \le n} f^{\oo_j \cap \oo_{j'} \cap \oo_{j''}}(\vx) - \ldots
\label{eq:in_ex_gen}
}
The time-$t$ \textbf{in-ex information} is then defined as
\eq{
&\II^{\NN^*}(\vx(t)) := \left[\widehat{\sum_{\oo \in {\NN^*}}}   s^\oo(\vx(t))\right] - s(\vx(t)) 
\label{eq:in_ex_info}
}
(A related concept, not involving a dependency structure, is called ``co-information'' in~\cite{bell2003co}.)
As an example, if ${\NN^*} $ consists of two units, 
$\oo_1, \oo_2$, with no intersection, then the expected in-ex information at time $t$ is just
the mutual information between those units at that time.
More generally, if there an arbitrary number of units in ${\NN^*}$
but none of them overlap, then the expected in-ex information is the 
``total correlation''~\cite{ting1962amount,wolpert2020minimal}, 
\eq{
I\left( X^{\oo_1}(t); X^{\oo_2}(t);  \ldots \right) &:=  \left(\sum_\oo \langle s^\oo(t) \rangle\right)  - \langle s(t) \rangle 
}

Since dependency structures are closed under intersections, if we apply the inclusion-exclusion principle
to the expressions for local and global EF
we get
\eq{
Q(\vx) 	&= \widehat{\sum_{\oo \in {\NN^*}}}  {Q}^\oo(\vx)
\label{eq:heat_decomp}
} 
Combining this with \cref{eq:43,eq:local_EP,eq:global_EP,eq:in_ex_info} gives
\eq{
\sigma^\NN(\vx) &=   \widehat{\sum_{\oo \in {\NN^*}}}  \sigma^{\oo}(\vx) - \Delta \II^{{\NN^*}}(\vx)
\label{eq:global_EP_decomp_in_ex}
}
Taking expectations of both sides of \cref{eq:global_EP_decomp_in_ex} we get
\eq{
\langle \sigma^\NN \rangle &=   \widehat{\sum_{\oo \in {\NN^*}}} \langle \sigma^{\oo}\rangle -\langle \Delta \II^{{\NN^*}}\rangle
\label{eq:29}
}
As a simple example,
if there are no overlaps between any units,
then expected global EP reduces to
\eq{
\langle \sigma^\NN \rangle &= \sum_\oo \langle \sigma^\oo \rangle -   \langle\Delta I\left( X^{\oo_1}; X^{\oo_2};  \ldots \right)  \rangle
		\\
	&\ge  -   \langle\Delta I\left( X^{\oo_1}; X^{\oo_2};  \ldots \right)  \rangle
\label{eq:in_ex_decomp}
}
which is a strengthened second law of thermodynamics.


\textit{Multipartite Fluctuation Theorems.--- } 
Write $\tilde{\vx}$ for $\vx$ reversed in time, i.e., $\tilde{ \vx}(t) = \vx(t_f - t)$.
Write $\vP(\vx)$  for the probability density function over trajectories
generated by {starting} from the initial distribution $p_x(0)$, and
write $\tilde{\vP}(\vx)$ for the probability density function over trajectories
generated by {starting} from the ending distribution $p_x(t_f)$, and then evolving 
according to the time-reversed sequence of rate matrices, $\tilde{K}(t) = K(t_f-t)$.

Let $\AAA$ be any set of units (not necessarily a dependency structure) and write
$\sigma^{ \AAA}(\vx)$ for the total EP generated by the systems in $\AAA$ under $\vP$.
(Note that since $\cup \AAA$ is a union of units, we can write $\sigma^{ \AAA}(\vx)$ as $\sigma^{ \AAA}(\vx_\AAA)$.)
Define $\vec{\sigma}^\AAA$ as the vector whose
components are the local EP values $\sigma^\alpha$ for all $\alpha \in \AAA$ (including $\alpha = \AAA$).
One can prove the associated detailed fluctuation theorem (DFT)
\eq{
\ln \left[\dfrac{\vP(\vec{\sigma}^\AAA)}{ \tilde{\vP}(-\vec{\sigma}^\AAA)}\right]  &= \sigma^{ \AAA}
\label{eq:50}
}
where $ \tilde{\vP}(-\vec{\sigma}^\AAA)$
is the joint probability of the specified vector of EP values under ${\tilde {\vP}}$.
%

Subtracting instances of \cref{eq:50} evaluated for different choices of $\AAA$ 
gives conditional DFTs, which in turn give conditional integral fluctuation theorems (IFTs). 
As an example, subtract \cref{eq:50} for $\AAA$ set to some singleton $\{\oo\}$ from 
\cref{eq:50} for $\AAA = \NN$. This gives
\eq{
\left\langle e^{\sigma^\oo - \sigma^\NN} \,\vert\,  \sigma^\oo \right\rangle &= 1
\label{eq:28}
}
(for any specific EP value $\sigma^\oo$ such that both $\vP(\sigma^\oo)$ and $\tilde{\vP}(-\sigma^\oo)$ are nonzero).
Jensen's inequality then gives $\left\langle \sigma^\NN \,\vert\, \sigma^\oo \right\rangle \ge \sigma^\oo$
and so averaging over $\sigma^\oo$,
\eq{
 \left\langle \sigma^\NN\right\rangle \ge \left\langle \sigma^\oo \right\rangle
\label{eq:38}
}
%
Plugging in \cref{eq:global_EP_decomp_in_ex} then gives
\eq{
{\widehat{\sum}}_{\oo'\in {\NN^*}} \langle \sigma^{\oo'}\rangle - \langle\Delta \II^{{\NN^*}}\rangle \ge 
 \langle \sigma^{\oo}\rangle
\label{eq:23d}
}

The \textbf{multi-divergence} for a set of units $\AAA$ is 
\eq{
		D\left(\vP(\vec{\sigma}^\AAA) \; ||\; \tilde{\vP}(-\vec{\sigma}^\AAA)\right) 
		- 	\sum_{\AAA_i \in \AAA} D\left(\vP({\sigma}^{\AAA_i}) \;||\; \tilde{\vP}(-\sigma^{\AAA}_i)\right)
}
and written as $I\left(\vP(\vec{\sigma}^\AAA) \; ||\; \tilde{\vP}(-\vec{\sigma}^\AAA)\right)$~\cite{wolpert_thermo_comp_review_2019}.
It is an extension of the concept of the total correlation of the systems in $\AAA$, replacing entropies with relative entropies 
between forward and backward trajectories. One can prove that
\eq{
\left\langle \sigma^\NN \right\rangle &\ge \sum_{i = 1}^m \left\langle \sigma^{\AAA_i}  \right\rangle + I\left(\vP(\vec{\sigma}^\AAA) \; ||\; 		
				\tilde{\vP}(-\vec{\sigma}^\AAA)\right) 
\label{eq:22}
}
with equality if $\cup_i \AAA = \NN$. Note that the sum of local EPs in \cref{eq:22} is a normal sum, not an in-ex sum. 

Often one can show that multi-divergence is non-negative, and
so $\left\langle \sigma^\NN \right\rangle \ge \sum_{i = 1}^m \left\langle \sigma^{\AAA_i} \right\rangle$.
As an example, multi-divergence reduces to total correlation (and so is non-negative) if
$\tilde{\vP}(-\vec{\sigma}^\AAA)$ is a product distribution. Moreover, often
$\tilde{\vP}(-\vec{\sigma}^\AAA)$ is close to being a product distribution 
in a relaxation process (see Supplemental Material at
[URL will be inserted by publisher]). 

\begin{figure*}[tbp]
\includegraphics[width=125mm]{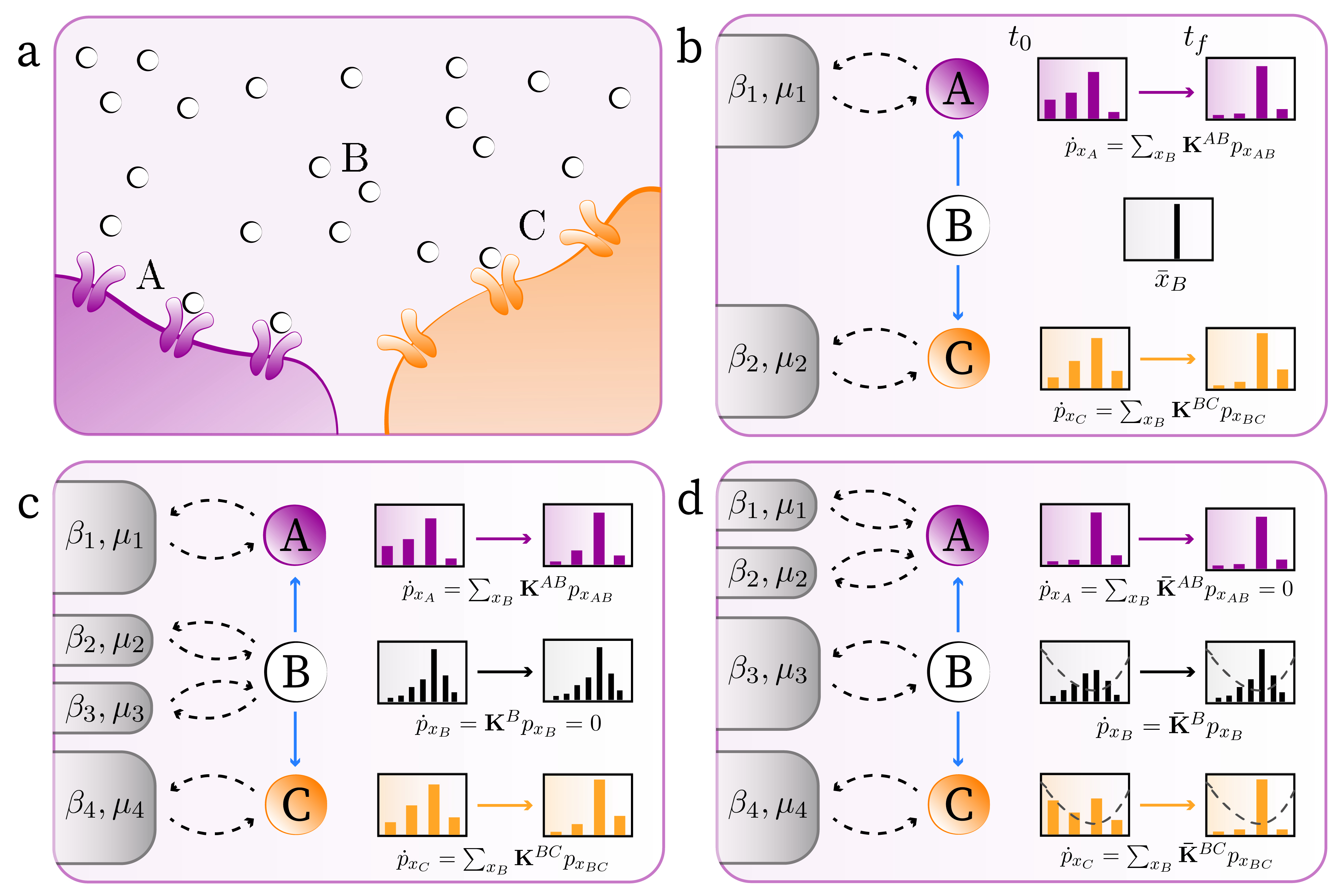}
\caption{
(a) The MPP discussed in the text.
The state of the intercellular medium (system $B$) is the (discretized) concentration of
a ligand and the states of the cell-wall receptors  (systems $A, C$)
are their respective observations of that concentration. (b) In this scenario system $B$ does not change its state.
(Histograms indicate starting and ending distributions.)
(c) In this scenario system $B$ is in an NESS. (d) In this scenario 
joint system $AB$ is in an NESS. (Overbars indicate
time-homogeneous rate matrices.)
}
\label{fig:2}
\end{figure*}

\textit{Extended TURs and strengthened second law.--- } 
The simplified ligand-sensing example in \cref{fig:2}
can be used to illustrate the consequences of these results.
To begin, evaluate \cref{eq:23d} for this example first for $\oo = AB$
and then for $\oo = BC$, to get
\eq{
\min \left[\left\langle \sigma^{BC} \right \rangle, \left\langle \sigma^{AB} \right \rangle \right]
			- \left\langle \sigma^B \right\rangle \ge   \Delta I(A; C \,|\, B)
\label{eq:23dd}
}
where $I\left(A; C \,|\, B\right)(t)$ is the conditional mutual
information at time $t$ between $x_A$ and $x_C$, given $x_B$.
\cref{eq:23dd} is a new type of speed limit theorem:
If one wants the process to (change the distributions in order
to) have large $\Delta I(A; C \,|\, B)$, one must pay for it with large value of both
$\left\langle \sigma^{AB} \right \rangle$ and $\left\langle \sigma^{BC} \right \rangle$. (Special cases of this
kind of speed limit were implicit in some of the results in~\cite{wolpert_thermo_comp_review_2019,wolpert_kolchinsk_first_circuits_published.2020,wolpert.thermo.bayes.nets.2020}.)

Next, suppose $x_B$ is constant during the process. Then
\eq{
\left \langle \sigma^\NN \right \rangle =  \left\langle \sigma^{AB} \right\rangle + \left\langle \sigma^{BC}\right\rangle 
  -  \Delta I(A; C \,|\, B)  \ge -   \Delta I(A; C \,|\, B) 
\label{eq:25}
}
by \cref{eq:global_EP_decomp_in_ex}.
One can prove
that $\Delta I(A; C \,|\, B) \le 0$ if $x_B$ is constant. So when the ligand concentration in the medium
is constant (to within the precision of the coarse-grained binning of $X_B$), 
\cref{eq:25} is stronger than the conventional second law (see \cref{fig:2}(b)).

Now suppose only that system $B$ is subject to a TUR, e.g., it's in an NESS, so that
$\langle \sigma^B \rangle \ge 2\langle J_B\rangle^2 / {\mbox{Var}}(J_B)$,
where $J_B(\vx_B)$ is an arbitrary current among $B$'s
states  (see \cref{fig:2}(c)).
Plugging this into \cref{eq:23dd}, applying \cref{eq:38} for $\oo = BC$, and then applying \cref{eq:38} 
for $\oo = B$, we derive a new kind of TUR:
\eq{
\langle \sigma^\NN\rangle \ge 2\langle J_B\rangle^2 / {\mbox{Var}}(J_B) + \max\left[ \Delta I(A; C \,|\, B), 0\right] 
\label{eq:27aa}
}

Next, choose $\AAA = \{AB, BC\}$ in  \cref{eq:22} to get
\eq{
\left\langle \sigma^\NN \right\rangle \ge \left\langle \sigma^{AB}  \right\rangle + \left\langle \sigma^{BC}  \right\rangle 
	+ I\left(\vP(\sigma^{AB}, \sigma^{BC}) \; ||\; \tilde{\vP}(-\sigma^{AB}, -\sigma^{BC})\right)
\label{eq:28}
}
One can prove
that for broad classes of MPPs,
$I\left(\vP(\sigma^{AB}, \sigma^{BC}) \; ||\; \tilde{\vP}(-\sigma^{AB}, -\sigma^{BC})\right) \ge 0$.
In particular this holds if the rate matrices are time-homogeneous, the Hamiltonian is uniform
and unchanging, and  $B$ relaxes to its equilibrium distribution by time $t_f$.
In this scenario the relaxing unit $BC$ obeys the conditions 
for the arbitrary initial state TUR~\cite{liu2020thermodynamic}. Suppose as well
that $AB$ obeys the conditions for the NESS-based TUR  (see \cref{fig:2}(d)).
Then \cref{eq:28} lower bounds expected global EP in terms of current precisions in \textit{different} units:
\eq{
\langle \sigma^\NN\rangle  &\ge  
2 \dfrac{\langle J_{AB}\rangle^2}{ {\mbox{Var}}(J_{AB})} + 
		\dfrac{\langle t_f j_{BC}(t_f)\rangle^2}{ {\mbox{Var}}(J_{BC})}
\label{eq:29aa}
}
($j_{BC}(t_f)$ is the derivative of the expectation of an arbitrary current in joint system $BC$, evaluated at $t_f$).
Importantly, \cref{eq:27aa,eq:29aa} hold even if the full system does not satisfy conditions for {any} conventional TUR.

Finally, combining \cref{eq:28,eq:29}
gives 
\eq{
 \left\langle \sigma^B \right\rangle \le  -\Delta I(A; C \,|\, B)
	- I\left(\vP(\sigma^{AB}, \sigma^{BC}) \; ||\; \tilde{\vP}(-\sigma^{AB}, -\sigma^{BC})\right)
\label{eq:30aa}
}
When $I\left(\vP(\sigma^{AB}, \sigma^{BC}) \; ||\; \tilde{\vP}(-\sigma^{AB}, -\sigma^{BC})\right) \ge 0$  (e.g., in \cref{fig:2}(d)), 
\cref{eq:30aa} \textit{upper}-bounds EP in the intercellular medium in terms of a purely
information-theoretic quantity, $-\Delta I(A; C \,|\, B)$, which also involves the cell wall receptor systems $A$ and $C$.
(\cref{eq:30aa}  also shows that it is impossible to have both $\Delta I(A; C \,|\, B) > 0$ and $I\left(\vP(\sigma^{AB}, \sigma^{BC}) \; ||\; \tilde{\vP}(-\sigma^{AB}, -\sigma^{BC})\right) > 0$.)

\textit{Discussion.--- } 
In  many sets of
co-evolving systems, the times of each system's state transitions is
random. 
In this paper I derive FTs that govern such sets of systems. 
I then use those FTs to derive a strengthened second law, and to
substantially extend all currently known TURs: even if the overall system does not meet any criteria for 
a TUR, the expected EP of the overall system is lower-bounded by the current precisions of
any of the constituent systems that separately meet such criteria.

$ $

\textit{Acknowledgements.--- } 
I thank Farita Tasnim, 
G{\"u}lce Kardes,
and 
Artemy Kolchinsky 
for helpful discussion. 
This work was supported by the Santa Fe Institute, US NSF
Grant CHE-1648973  and FQXi Grant  FQXi-RFP-IPW-1912.

\bibliographystyle{amsplain}
\bibliography{../../../../LANDAUER.Shared.2016/thermo_refs.main,../../../../LANDAUER.Shared.2016/thermo_refs_2}


\appendix

\section{Entropy flow into systems and units}
\label{app:def_EF_units}

To fully define the entropy flow into systems we need to introduce some more notation.
Let $M(\vx)$ be the total number of state transitions during the time interval $[0,t_f]$ by all systems 
(which might equal $0$). If $M(\vx) \ge 1$, define $\eta_\vx : \{1, \ldots, M(\vx)\} \rightarrow \NN$ as the function that maps
any integer $j \in  \{1, \ldots, M(\vx)\}$ to the system that changes its state in the $j$'th transition.  Let $k(j)$ be the
associated function specifying which reservoir is involved in that $j$'th transition. (So for all $j$, $k(j)$ specifies
a reservoir of system $\eta(j)$.) Similarly,
let $\tau_\vx : \{0, \ldots, M(\vx)\} \rightarrow \NN$ be the function that maps
any integer $j \in  \{1, \ldots, M(\vx)\}$ to the time of the $j$'th transition, and maps $0$ to the time $0$. 

From now on,
I leave the subscript $\vx$ on the maps $\eta_\vx$ and $\tau_\vx$ implicit.
So for example, $\eta^{-1}(i)$ is the set (of indices specifying) all state transitions at which system $i$ changes state in the trajectory $\vx$. 
More generally, for any set of systems $\alpha$, $\eta^{-1}(\alpha) := \cup_{i \in \alpha} \eta^{-1}(i)$ is the set
of  all state transitions at which a system $i \in \alpha$ changes state in the trajectory $\vx$. 

Given these definitions, the total entropy flow into system $i$ from its reservoirs during $[0,t_f]$ is
\eq{
Q^i(\vx) &:=  \sum_{j \in \eta^{-1}(i)}  
\beta_i^{k(j)}\bigg(H_{{\vx}_{r(i)}(\tau(j))}(i;\tau(j)) - H_{{\vx}_{r(i)}(\tau(j-1))}(i;\tau(j))  \nonumber \\
	&\qquad + \mu_{i}^{k(j)}\left[n_i^{k(j)}({\vx}_i(\tau(j-1)) - n_i^{k(j)}({\vx}_i(\tau(j))\right]\bigg) 
\label{eq:19a}
}
where I interpret the sum on the RHS to be zero if system $i$ never undergoes a state transition in trajectory $\vx$.
As mentioned in the text, the local EF into a unit $\alpha$ for trajectory $\vx$ is
just the sum of the EFs into all the systems in $\alpha$ for that trajectory.
%

Expanding, under \cref{eq:nldb},
\eq{
Q^\alpha(\vx) &= \sum_{i \in \alpha}	\sum_{j \in \eta^{-1}(i)}  
						\ln	\left[\dfrac{K^{{\vx}_{r(i)}(\tau(j))}_{{\vx}_{r(i)}(\tau(j-1))}(i; k(j), \tau(j))}  {K^{{\vx}_{r(i)}(\tau(j-1))}_{{\vx}_{r(i)}(\tau(j))}(i; k(j), \tau(j))}\right]
\label{eq:local_EF}
}
So in the special case that $\alpha$ is a unit, 
\eq{
Q^\alpha(\vx) &= \sum_{i \in \alpha}	\sum_{j \in \eta^{-1}(i)}  
						\ln	\left[\dfrac{K^{{\vx}_{\alpha}(\tau(j))}_{{\vx}_{\alpha}(\tau(j-1))}(i; k(j), \tau(j))}  {K^{{\vx}_{\alpha}(\tau(j-1))}_{{\vx}_{\alpha}(\tau(j))}(i; k(j), \tau(j))}\right]
\label{eq:local_EF_unit}
}
Note that ${\vx}_{-i}(\tau(j-1)) = {\vx}_{-i}(\tau(j))$ for all systems $i$, for all $j \in \eta^{-1}(i)$, 
since the process is multipartite. Therefore the global EF can be written as
\eq{
Q(\vx) = \sum_i Q^i(\vx) 
	= \sum_{j =1}^{M(\vx)}  
						\ln	\left[\dfrac{K^{{\vx}(\tau(j))}_{{\vx}(\tau(j-1))}(\tau(j))}  {K^{{\vx}(\tau(j-1))}_{{\vx}(\tau(j))}(\tau(j))}\right]
\label{eq:30}
}

As a final, technical point, note that $\vx$ is fully specified by $M, \tau, k, \eta$ and the finite
list of the precise state transitions at the times listed in $\tau$ by the associated systems listed in $\eta$
mediated by the associated reservoirs listed in $k$. So the probability measure of a trajectory $\vx$ is 
\eq{
\vP(\vx) = P(M) P(\vx \,|\, M, \tau, k, \eta) P(\tau, k, \eta \,|\, M)
}
For each integer $M$, all the terms on the RHS are either probability distributions or probability density functions,
and therefore we can define the integral over $M, \tau, k, \eta, \vx$.
So in particular, 
$\delta$ function over trajectories in the equations in the main text is shorthand for a 
function that equals zero everywhere that its argument is nonzero, and such that its
integral equals $1$.

\section{Example where global Hamiltonian cannot equal sum of local Hamiltonians under SLDB}
\label{app:two-way-observation}

The global Hamiltonian need not be a sum of the local Hamiltonians.
Indeed, it's possible that all local Hamiltonians equal one another and also equal the global Hamiltonian.
To see this, 
consider a scenario where there are exactly two distinct systems $i, j$, which both observe one another very closely, so that the rate matrices
of both of them have approximately as much dependence on the state of the \textit{other} system as on their own state.
In this scenario, there is a single unit, $r(i) = r(j) = \{i, j\} = \NN$. Assume as well that each
system is only connected to a single heat bath, with no other reservoirs.

If we took the global Hamiltonian to be
$H_x(t) = H_x(i; t) + H_x(j; t)$, its change under the fluctuation $(x'_i, x_{j}) \rightarrow (x_i, x_{j})$ would be
\eq{
H_{x_i, x_{j}}(i;t) - H_{x'_i, x_{j}}(i;t) + H_{x_i, x_{j}}(j;t) - H_{x'_i, x_{j}}(j;t)
\label{eq:d6}
}
This is the change in energy of the total system during the transition. Moreover, since this state transition 
only changes the state of system $i$ and since we assume the dynamics is a MPP, only the heat bath of system $i$ could 
have changed its energy during the transition. Therefore by conservation of energy,
the change in the energy of the bath of system $i$ must equal the expression
in \cref{eq:d6}. 

Under SLDB, that change in the energy
of the heat bath of system $i$ would be the change in the local Hamiltonian
of system $i$. 
Therefore to have SLDB be even approximately true, it would have to be the case that
\eq{
|H_{x_i, x_{j}}(i;t) - H_{x'_i, x_{j}}(i;t)|  \;&\gg\;   |H_{x_i, x_{j}}(j;t) - H_{x'_i, x_{j}}(j;t)|
\label{eq:d7}
}
for all  pairs $x, x'$.
%
Similarly, to address the case where a fluctuation to $x_j$ arises due to $j$'s interaction
with \textit{its} heat bath, we would need to have
\eq{
|H_{x_i, x_{j}}(j;t) - H_{x_i, x'_{j}}(j;t)| \;&\gg\;  |H_{x_j, x_{j}}(i;t) - H_{x_i, x'_{j}}(i;t)|
\label{eq:d8}
}
for all pairs $x, x'$.

Next, we need to formalize our requirement that ``the rate matrices
of both \{systems\} have approximately as much dependence on the state of the {other} system as on their own state''.
One way to do that is to require that the dependence on $x_j$ of (the ratio of forward and backward terms in) $i$'s rate matrix is 
not much smaller than the typical terms in that rate matrix, i.e., for all pairs $x, x'$,
\eq{
&\left| \left(H_{x_i, x_{j}}(i;t) - H_{x'_i, x_{j}}(i;t)\right)  - \left(H_{x_i, x'_{j}}(i; t) - H_{x'_i, x'_{j}}(i; t)\right) \right| \nonumber \\
	&\qquad \widesim \;
\dfrac{\left| H_{x_i, x_{j}}(i;t) - H_{x'_i, x_{j}}(i;t)\right|  + \left| H_{x_i, x'_{j}}(i; t) - H_{x'_i, x'_{j}}(i; t) \right| }{2}  \nonumber \\
	&\qquad := \Delta_{i;x,x'}
\label{eq:d9}
}
Similarly, since system $j$ is also observing $i$,
\eq{
&\left| \left(H_{x_i, x_{j}}(j;t) - H_{x_i, x'_{j}}(j;t)\right)  - \left(H_{x'_i, x_{j}}(j; t) - H_{x'_i, x'_{j}}(j; t)\right) \right| \nonumber \\
	&\qquad \widesim \;
\dfrac{ \left| \left(H_{x_i, x_{j}}(j;t) - H_{x_i, x'_{j}}(j;t)\right)  + \left(H_{x'_i, x_{j}}(j; t) - H_{x'_i, x'_{j}}(j; t)\right) \right| }{2}   \nonumber \\
	&\qquad := \Delta_{j;x,x'}
\label{eq:d10}
}

Plugging \cref{eq:d7} into the definition of $\Delta_{i; x, x'}$ and then using the fact that for any real numbers $a, b$, $|a - b| < |a| +|b|$, we get
\eq{
&\left| \left(H_{x_i, x_{j}}(j;t) - H_{x'_i, x_{j}}(j;t)\right)  - \left(H_{x_i, x'_{j}}(j; t) - H_{x'_i, x'_{j}}(j; t)\right) \right| \nonumber \\
	&\qquad \ll \Delta_{i;x,x'}
\label{eq:d11}
}
In addition, shuffling terms in the LHS of \cref{eq:d9} gives
\eq{
&\left| \left(H_{x_i, x_{j}}(i;t) - H_{x_i, x'_{j}}(i; t) \right)  - \left(H_{x'_i, x_{j}}(i;t)  - H_{x'_i, x'_{j}}(i; t)\right) \right| \nonumber \\
	&\qquad \widesim \;  \Delta_{i; x,x'}
\label{eq:d12}
}

The analogous reasoning with \cref{eq:d8,eq:d10} gives
\eq{
&\left| \left(H_{x_i, x'_{j}}(i;t) - H_{x_i, x_{j}}(i;t)\right)  - \left(H_{x'_i, x'_{j}}(i; t) - H_{x'_i, x_{j}}(i; t)\right) \right| \nonumber \\
	&\qquad \ll \Delta_{j;x,x'}
\label{eq:d13}
}
and
\eq{
&\left| \left(H_{x_i, x_{j}}(j;t) - H_{x'_i, x_{j}}(j; t) \right)  - \left(H_{x_i, x'_{j}}(j;t)  - H_{x'_i, x'_{j}}(j; t)\right) \right| \nonumber \\
	&\qquad \widesim \;  \Delta_{j; x, x'}
\label{eq:d14}
}

Comparing \cref{eq:d12,eq:d13} establishes that $\Delta_{i; x, x'} \ll \Delta_{j; x, x'}$. However,
comparing \cref{eq:d11,eq:d14} establishes that $\Delta_{j; x, x'} \ll \Delta_{i; x, x'}$. This contradiction shows
that it is not possible for the global Hamiltonian to be a sum of the two local Hamiltonians, given that
both systems ``observe one another very closely''. (See \cref{app:exact_ldb}.)

\section{Alternative expansion of trajectory-level global EP }
\label{app:chi_decomp}

\subsection{Definition of $\chi^\oo(\vx)$}
First, for any unit $\oo$, define 
\eq{
\chi^\oo(\vx) := \sigma^\NN(\vx) - \sigma^\oo(\vx)
\label{eq:chi_def}
}
I now show that 
\eq{
 \chi^\oo(\vx) 	&= \Delta s^{-\oo \,|\, \oo}(\vx) - Q^{-\oo}(\vx)
\label{eq:global_EP_decomp_windowing}
}
where
\eq{
\Delta s^{-\oo \,|\, \oo}(\vx) &= \Delta s^{-\oo, \oo}(\vx) - \Delta s^{\oo}(\vx) \\
	&= -\Delta \ln p_{\oo,-\oo}(\vx) +  \Delta \ln p_{\oo}(\vx) \\
	&=  -\Delta \ln p_{\NN}(\vx) +  \Delta \ln p_{\oo}(\vx)
}
is the change in the value of the conditional stochastic entropy of the entire system given the state of $\oo$.
Note that in general $-\oo$  will not be a unit. So the
entropy flow into the associated reservoirs, $Q^{-\oo}(\vx)$, may depend on the trajectory of systems outside of $-\oo$,
i.e., it may depend on $\vx_{\oo}$. 

As shorthand, from now on I leave the function $\tau$ implicit, so that for example, $\vx(\tau(j))$ gets shortened
to $\vx(j)$. (Note though that with slight abuse of notation, I still take $\vx(t_f)$ to mean the state of the
system at $t = t_f$ under trajectory $\vx$.)
Given any unit $\oo$, we can expand the global EP as


\begin{widetext}

\eq{
\sigma^\NN(\vx) &=   \ln p_{\vx(0)}(0) - \ln p_{\vx(t_f)}(t_f) + \sum_{j =1}^{M(\vx)}  
						\ln	\left[\dfrac{K^{{\vx}(j-1)}_{{\vx}(j)}(\eta(j);j)}  {K^{{\vx}(j)}_{{\vx}(j-1)}(\eta(j);j)}\right]  
 																		\nonumber \\
	&=  \ln p^{X_\oo}_{\vx_\oo(0)}(0)) - \ln p^{X_\oo}_{\vx_\oo(t_f)}(t_f) + \sum_{j \in \eta^{-1}(\oo)} 
		\ln \left[\dfrac{K^{{\vx_\oo}(j-1)}_{{\vx_\oo}(j)}(\eta(j);j)} 
				 {K^{{\vx}_\oo(j)}_{{\vx_\oo}(j-1)}(\eta(j);j) }\right]   \nonumber \\
	&	\qquad\qquad+  \ln p^{X|X_\oo}_{\vx(0)}(0)  - \ln p^{X|X_\oo}_{\vx(t_f)}(t_f) 
				  + \sum_{j \in \eta^{-1}(-\oo)} 	\ln \left[\dfrac{K^{{\vx}(j-1)}_{{\vx}(j)}(\eta(j);j)}  				
				{K^{{\vx}(j)}_{{\vx}(j-1)}(\eta(j);j) }\right] 			\nonumber \\
	&= \sigma^\oo(\vx) + \ln p^{X|X_\oo}_{\vx(0)}(0)  - \ln p^{X|X_\oo}_{\vx(t_f)}(t_f)
				  + \sum_{j \in \eta^{-1}(-\oo)} 	\ln \left[\dfrac{K^{{\vx}(j-1)}_{{\vx}(j)}(\eta(j);j)}  				
				{K^{{\vx}(j)}_{{\vx}(j-1)}(\eta(j);j) }\right] 
\label{eq:27b}
}
\end{widetext}

Expand
\eq{
\Delta s^{{-\oo} |\oo}(\vx) &= \Delta s^{(-\oo,\oo) \,|\, \oo}(\vx)  \\
	&= \Delta s^{\NN \,|\, \oo}(\vx)  \\
		&=  \ln p^{\NN|\oo}_{\vx(0)}(0)  - \ln p^{\NN|\oo}_{\vx(t_f)}(t_f) 
}
This allows us to rewrite \cref{eq:27b} more succinctly as
\eq{
\chi^\oo(\vx) &=  \Delta s^{\NN \,|\, \oo}(\vx) 
		 + \sum_{j \in \eta^{-1}(-\oo)}	\ln \left[\dfrac{K^{{\vx}(j-1)}_{{\vx}(j)}(\eta(j);j)}  				
				{K^{{\vx}(j)}_{{\vx}(j-1)}(\eta(j);j) }\right] \nonumber \\
	&= \Delta s^{\NN \,|\, \oo}(\vx) - Q^{-\oo}(\vx)
\label{eq:27a}
}
which establishes \cref{eq:global_EP_decomp_windowing}.

\subsection{FTs involving $\chi^\oo(\vx)$}

To gain insight into \cref{eq:27a}, define the counterfactual rate matrix $\underline{K}(t) := K(-\oo; t)$,
and let $\underline{\vP}$ be what the density over trajectories $\vx$
would have been if the system had evolved from the initial distribution $p_x(0)$ under $\underline{K}(t)$ rather than $K(t)$. 
Define $\Delta s^{{-\oo}}_{\underline{\vP}}(\vx)$ and $\sigma_{\underline{\vP}}(\vx)$ accordingly.
Then we can expand the second term on the RHS of \cref{eq:27a} as
\eq{
Q^{-\oo}(\vx) &= \sigma_{\underline{\vP}}(\vx) - \Delta s^{{-\oo}}_{\underline{\vP}}(\vx)
\label{eq:33maintext}
}
So the heat flow from the baths connected to $-\oo$ into the associated systems
is the difference between a (counterfactual) global EP
and a (counterfactual) change in the entropy of those systems.

We can iterate these results, to get more refined decompositions of global EP. For example,
let $\underline{\NN}^*$ be a dependency structure of $\underline{K}$, the counterfactual
rate matrix defined just before \cref{eq:33maintext}.
Let $\oo$ be a unit in ${\NN^*}$ while
$\alpha$ is a unit in $\underline{\NN}^*$. Then we can insert \cref{eq:33maintext} 
into \cref{eq:global_EP_decomp_windowing} and apply that  \cref{eq:global_EP_decomp_windowing}  again,
to the resulting term $\sigma_{\underline{\vP}}$, to get
\eq{
\sigma^\NN(\vx)  &= \sigma^\oo(\vx)  + \left[\sigma^\alpha_{\underline{\vP}}(\vx) + \chi^{\alpha}_{\underline{\vP}}(\vx)
			+ \Delta s^{\NN \,|\, \oo}_\vP(\vx) - \Delta s^{{-\oo}}_{\underline{\vP}}(\vx)
\right]
\label{eq:c12}
}

Note that in general, $\alpha$ might contain systems outside of $\NN\setminus \oo$. As a result,
it need not be a unit of the full rate matrix $K$.
In addition, both (counterfactual) rates
$d \left\langle \sigma^\alpha_{\underline{\vP}} \right\rangle_{\underline{\vP}} /dt$
and  $d \left\langle \chi^\alpha_{\underline{\vP}} \right\rangle_{\underline{\vP}} /dt$ are non-negative.
However, if we evaluate those two rates under the actual density $\vP$ rather than
the counterfactual $\underline{\vP}$, it may be that one or the other of them is negative. This is just like how the expected values of
the analogous ``EP'' terms in~\cite{ito2013information,shiraishi2015fluctuation}, which concern a single
system, may have negative derivatives.

The trajectory-level decomposition of global EP in terms of $\chi^\oo$ can be exploited to get
other FTs in addition to those presented in the main text. For example,
if we plug \cref{eq:chi_def} into \cref{eq:28} instead of plugging in \cref{eq:global_EP_decomp_in_ex}, 
and then use  \cref{eq:global_EP_decomp_windowing},  we get
\eq{
\left\langle e^{- \chi^\oo}  \,|\, \sigma^\oo \right\rangle &= 
 \left\langle \exp \left( Q^{-\oo} - \Delta s^{\NN|\oo}\right) \;\bigg| \; \sigma^\oo \right\rangle  = 1 
\label{eq:32c}
}
In contrast, the analogous expression using the in-ex sum expansion in the main text is
\eq{
\left\langle e^{\left(\Delta \II^{{\NN^*}} + \sigma^\oo - {\widehat{\sum}}_{\oo'\in {\NN^*}} \sigma^{\oo'} \right)}
			 \;\bigg| \; \sigma^\oo  \right\rangle = 1   
\label{eq:32b}
}
Applying Jensen's inequality to \cref{eq:32b} allows us to bound the change in the conditional entropy
of all systems \textit{outside} of any unit $\oo$, given the joint state of the systems
within $\oo$, by the EF of those systems:
\eq{
\Delta S^{\NN \;|\; \oo} \ge Q^{\NN \setminus \oo}
} 
(\cref{eq:32c} can often be refined by mixing and matching among alternative expansions of $\chi^\oo$,
along the lines of \cref{eq:c12}.)

Applying Jensen's inequality to \cref{eq:32c}  shows that for all $\sigma^\oo$ with nonzero probability,
\eq{
\left\langle \Delta s^{\NN|\oo} - Q^{-\oo} \,\bigg|\, \sigma^\oo \right\rangle &\ge 0  
\label{eq:39aa}
}
Averaging both sides of \cref{eq:39aa} over all $\sigma^\oo$ gives
\eq{
 \Delta S^{\NN|\oo} &\ge  \left\langle Q^{-\oo} \right\rangle
\label{eq:40}
}
So  the expected heat flow into the systems outside of $\oo$ during any interval is upper-bounded
by the change during that interval in the value of the conditional entropy of the full system given the entropy of the unit $\oo$.

Since EPs, $\chi$'s, etc., all go to $0$ as $t_f \rightarrow 0$,
these bounds can all be translated into bounds concerning time derivatives. For example, in~\cite{wolpert2020minimal} it is
shown that $d \left\langle \chi^\oo \right\rangle /dt \ge 0$. Applying Jensen's inequality
to \cref{eq:32c} gives a strengthened version of this result: for any value
of $\sigma^\oo$ that has nonzero probability throughout an interval $t \in [0, t'>0]$, 
$d \left\langle \chi^\oo \,|\, \sigma^\oo \right\rangle / dt  \ge 0$ at $t=0$. 

%

\subsection{Rate of change of expected $\chi^\oo(\vx)$}

I now show that $d \left\langle \chi^\oo \right \rangle /dt$ is  
the sum of two terms. The first term is the expected global EP rate under a counterfactual rate matrix.
The second term is (negative of) the derivative of the mutual information between $X_\oo$ and $X_{-\oo}$, under a counterfactual
rate matrix in which $x_{-\oo}$ never changes its state. 
(This second term is an extension of what is sometimes called the ``learning rate'' in~\cite{wolpert2020minimal,barato_efficiency_2014,hartich_stochastic_2014,hartich_sensory_2016,matsumoto2018role,Brittain_2017}
and is related to what is called ``information flow'' in~\cite{horowitz2014thermodynamics}.)
Both of these terms are non-negative. Plugging into \cref{eq:chi_def} then confirms that the expected EP
of the full system is lower-bounded by the expected EP of any one of its constituent units, $\oo$, a result
first derived in~\cite{wolpert2020minimal}.

As shorthand replace $t_f$ with $t$, and then expand 
\eq{
\left\langle  -\ln p^{X|X_\oo}_{\vx(t)}(t) \right \rangle &= -\sum_x p_x(t) \left(\ln p_x(t) - \ln p_{x_\oo}(t)\right)
}
Therefore,
\eq{
\dfrac{d}{dt} \left\langle - \ln p^{X|X_\oo}_{\vx(t)}(t) \right \rangle &= 
			-\sum_{x,x'} K^{x'}_x(t) p_{x'}(t) \ln p_x(t) \nonumber \\
		&\; + 	\sum_{x_\oo,x'_\oo} K^{x'_\oo}_{x_\oo}(\oo;t) p_{x'_\oo}(t) \ln p_{x_\oo}(t)
\label{eq:30aaa}
}

In addition, the sum in \cref{eq:27a} is just the total heat flow \textit{from} the systems in $-\oo$ \textit{into} 
their respective heat baths, during the interval $[0, t]$, if the system follows trajectory $\vx$. Therefore 
the derivative with respect to $t$ of the expectation of that sum is just the expected heat flow rate at $t$
from those systems into their baths,
\eq{
-\sum_{x,x'} K^{x'}_x(-\oo; t) p_{x'}(t)\ln \left[\dfrac{K^{x}_{x'}(-\oo;t) } {K^{x'}_{x}(-\oo;t) }\right]  
\label{eq:31a}
}

Note as well that $K(t) = K(\oo;t) + K(-\oo;t)$. So if we add \cref{eq:31a} to \cref{eq:30aaa},
and use the fact that rate matrices are normalized, we get
\eq{
\dfrac{d \left\langle \chi^\oo \right\rangle}{dt} &= -\sum_{x,x'} K^{x'}_x(\oo;t) p_{x'}(t) \ln p_{x|x_\oo}(t) \nonumber \\
		&\qquad + \sum_{x,x'} K^{x'}_x(-\oo; t)p_{x'}(t) \ln \left[\dfrac{K^{x'}_x(-\oo;t) p_{x'}(t)}{K^{x}_{x'}(-\oo;t) p_{x}(t)}\right]
\label{eq:31aa}
}

The first sum in \cref{eq:31aa} is called the ``windowed derivative'',
$\dfrac{d^\oo}{dt} S^{X|X_\oo}(t)$, in~\cite{wolpert2020minimal}. 
Since $\oo$ is a unit, it is the (negative) of the derivative of the mutual
information between $X_\oo$ and $X_{-\oo}$, under a counterfactual rate matrix in which $x_{-\oo}$ is held
fixed. As discussed in~\cite{wolpert2020minimal}, by the data-processing inequality, this term is non-negative.

The second sum in \cref{eq:31aa} is what was called $\left\langle \dot{\sigma^\NN}_{K(\NN\setminus \oo; t)} \right\rangle$ 
in~\cite{wolpert2020minimal}. Since it is the expected rate of EP for a properly normalized, 
counterfactual rate matrix, it too is non-negative. Therefore the full expectation $\dfrac{d \left\langle \chi^\oo \right\rangle}{dt}$
is non-decreasing in time.

This decomposition of $d \left \langle \sigma^\NN - \sigma^\oo \right \rangle / dt$ was first derived in~\cite{wolpert2020minimal}.
However, that derivation did not start from a trajectory-level definition of local and global EPs, as done here.

\section{Proof of vector-valued DFT}
\label{app:DFT_vector}
Proceeding in the usual way~\cite{van2015ensemble,esposito.harbola.2007,seifert2012stochastic}, we first want to calculate
\eq{
\ln \left[\dfrac{\vP(\vx_{\AAA})} {\tilde{\vP}(\tilde{\vx}_{\AAA})} \right]
\label{eq:63}
}
Paralleling the development in 
App.\,A of~\cite{esposito.harbola.2007}, we reduce this expression
to a sum of two terms. The first term is a sum, over all transitions in $\vx_\AA$, of
the log of the ratio of two associated entries in the rate matrix of the system that changes state in that transition
\footnote{If there are no chemical reservoirs, then since each system is coupled to its own heat bath,  
we can uniquely identify which bath was involved in each state transition in any given $\vx$ directly from $\vx$ itself.
This is not the case for trajectory-level analyses of systems which are
coupled to multiple mechanisms, e.g.,~\cite{esposito.harbola.2007}; to 
identify what bath is involved in each transition in that setting we need to know more than just $\vx$.}. 
Since a union of units is a unit, we can use \cref{eq:local_EF_unit} to show that
that first sum equals $-Q^{\AAA}(\vx_{\AAA})$. The second term is just $\Delta s^{\AAA}(\vx_{\AAA})$. 
Therefore by the definition of local EP, 
we have a DFT over trajectories, 
\eq{
\ln \left[\dfrac{\vP(\vx_{\AAA})} {\tilde{\vP}(\tilde{\vx}_{\AAA})} \right] &= \sigma^{\AAA}(\vx_{\AAA})
\label{eq:31}
}
where
\eq{
 \tilde{\vP}(-\vec{\sigma}^\AAA) &:= 
\int \mathcal{D}\tilde{\vx}_{\cup \AAA}\, \tilde{ {\vP}}(\tilde{ \vx}_{\cup\AAA}) 
		\prod_{\oo \in \AAA} \delta\left(-\sigma^\oo - \ln \left[\dfrac{\tilde{\vP}(\tilde{\vx}_\oo)}{{ {\vP}}({ \vx}_\oo)}\right]\right)
}
is the joint probability of the specified vector of EP values under ${\tilde {\vP}}$.

Similarly, for any single unit $\oo \in \AAA$,
\eq{
\ln \left[\dfrac{\vP(\vx_\oo)}{\tilde{\vP}(\tilde{\vx}_\oo)}\right] &= \sigma^\oo(\vx_\oo)
\label{eq:40b}
}
Therefore,
paralleling~\cite{van2015ensemble}, we can combine \cref{eq:31,eq:40b} to get a DFT for the probability density function of values of
$\vec{\sigma}^\AAA(\vx_{\AAA})$:
\eq{
\vP(\vec{\sigma}^\AAA) &= \int \mathcal{D}\vx_{\AAA} \, \vP(\vx_{\AAA}) 
		\prod_{\oo \in \AAA} \delta\left(\sigma^\oo - \ln \left[\dfrac{\vP(\vx_\oo)}{\tilde{ {\vP}}(\tilde{ \vx}_\oo)}\right]\right) \\
	& = e^{\sigma^{\AAA}} \int \mathcal{D}\vx_{\AAA} \, \tilde{ {\vP}}(\tilde{ \vx}_{\AAA}) 
		\prod_{\oo \in \AAA} \delta\left(\sigma^\oo - \ln \left[\dfrac{\vP(\vx_\oo)}{\tilde{ {\vP}}(\tilde{ \vx}_\oo)}\right]\right) \\
	& = e^{\sigma^{\AAA}} \int \mathcal{D}\tilde{\vx}_{\AAA}\, \tilde{ {\vP}}(\tilde{ \vx}_{\AAA}) 
		\prod_{\oo \in \AAA} \delta\left(-\sigma^\oo - \ln \left[\dfrac{\tilde{\vP}(\tilde{\vx}_\oo)}{{ {\vP}}({ \vx}_\oo)}\right]\right) \\
	& := e^{\sigma^{\AAA}} \tilde{\vP}(-\vec{\sigma^\NN})
}
i.e.,
\eq{
\ln \left[\dfrac{\vP(\vec{\sigma}^\AAA)}{ \tilde{\vP}(-\vec{\sigma}^\AAA)}\right]  &= \sigma^{\AAA}
\label{eq:50_app}
}
which establishes the claim. As always, the reader should note that
${ \tilde{\vP}(-\vec{\sigma}^\AAA)}$ is the probability of a reverse trajectory $\tilde{ \vx}$, generated under $\tilde{ {\vP}}$,
such that if \textit{its} inverse, $\vx$, had been generated in the forward process, it would have resulted in 
$\vec{\sigma}^\AAA$. That vector of EPs is the vector of EPs of the trajectory as measured using the formula for \textit{forward}-process EPs~\cite{van2015ensemble}. So in particular, it is the EPs using the drop in stochastic
entropy of the forward process, \textit{not} of the reverse process.

As an aside, note ``vector-valued fluctuation theorems'' were derived previously in~\cite{garcia2010unifying,garcia2012joint}, using
similar reasoning. However, those FTs did not involve vectors of local EPs. Rather than held for any choice of vector $(A_1(\vx,K(t)), A_2(\vx,K(t)), \ldots)$
such that: i) global EP of any trajectory is $\sigma^\NN(\vx) = \sum_i A_i(\vx,K)$; ii) each component
$A_i(\vx, K(t))$ is odd if you time-reverse the trajectory $\vx$ and the sequence of rate matrices, $K(t)$.
The first property does not hold in general for the vector of local EPs.

\section{Implications of the vector-valued DFT}
\label{app:other_implications}


In this appendix I derive \cref{eq:22} in the main text. I then present a simple example of the
claim made in the figure caption in the main text
that  for broad classes of MPPs, including the one depicted here,
$I\left(\vP(\sigma^{AB}, \sigma^{BC}) \; ||\; \tilde{\vP}(-\sigma^{AB}, -\sigma^{BC})\right) \ge 0$.
I follow this by briefly discussing some other implications of the
vector-valued DFT.


\subsection{Decomposition of expected global EP involving multi-divergence}

As in the main text, let $\AAA$ be any set of units, not necessarily a full unit structure.
Since $\sigma^{ \AAA}$ is a single-valued function of $\vec{\sigma}^\AAA$,
taking the average of both sides of the vector-valued DFT, \cref{eq:50}, over all $\vec{\sigma}^\AAA$ establishes that
\eq{
\left \langle \sigma^{ \AAA} \right \rangle &= D\left(\vP(\vec{\sigma}^\AAA) \,||\, \tilde{\vP}(-\vec{\sigma}^\AAA) \right)
\label{eq:40a}
}
(This is in addition to the fact that 
$\left \langle \sigma^\NN \right \rangle = D\left(\vP({\sigma^\NN}) \,||\, \tilde{\vP}(-{\sigma^\NN}) \right)$.)

If we now add and subtract $\sum_i \left\langle \sigma^{\AAA_i} \right\rangle$ on the RHS, 
we derive \cref{eq:22} in the  main text:
\eq{
\left\langle \sigma^\NN \right\rangle &\ge \sum_{i} \left\langle \sigma^{\AAA_i}  \right\rangle 
		+ D\left(\vP(\vec{\sigma}^\AAA) \; ||\; \tilde{\vP}(-\vec{\sigma}^\AAA)\right)   \nonumber \\
&\qquad\qquad 	- 	\sum_{i} D\left(\vP({\sigma}^{\AAA_i}) \;||\; \tilde{\vP}(-\sigma^{\AAA}_i)\right) \\
&=  \sum_{i} \left\langle \sigma^{\AAA_i}  \right\rangle + 
		I\left(\vP(\vec{\sigma}^\AAA) \; ||\; \tilde{\vP}(-\vec{\sigma}^\AAA)\right))
\label{eq:e6}
}
Intuitively, the multi-divergence on the RHS of \cref{eq:e6} measures how much of the distance
between $\vP(\vec{\sigma}^\AAA)$ and  $\tilde{ {\vP}}(-\vec{\sigma}^\AAA)$ arises from the correlations
among the variables $\{\sigma^{\AAA_i}\}$, in addition to the contribution from the marginal distributions of each variable
considered separately.

%
%
%
%

\subsection{Example of non-negativity of multi-divergence}

To illustrate how to calculate multi-divergence, use the chain rule for relative entropy to expand 
the multi-divergence in the figure caption 
in the main text as
\eq{
&I\left(\vP(\sigma^{AB}, \sigma^{BC}) \; ||\; \tilde{\vP}(-\sigma^{AB}, -\sigma^{BC} )\right) =   \nonumber \\
& \qquad D\left(\vP({\sigma}^{AB} \,|\, \sigma^{BC}) \; ||\; { \tilde{\vP}}(-{\sigma}^{AB} \,|\, -\sigma^{BC})\right) - 
		D\left(\vP({\sigma}^{AB}) \; ||\;  \tilde{\vP}(-{\sigma}^{AB})\right) 
\label{eq:e8a}
}
In words, this is the gain in how sure we are that a given value of $\sigma^{AB}$ was generated in the
forward trajectory rather than that  $-\sigma^{AB}$ was generated in the backward one, if we also know 
two other quantities: the value of $\sigma^{BC}$ when $\sigma^{AB}$ was
generated under the forward trajectory, and the value of  $-\sigma^{BC}$ when $-\sigma^{AB}$ 
was generated under the reverse trajectory.

Consider scenarios meeting the following conditions:
\begin{enumerate}
\item The rate matrices are time-homogeneous. 
\item The Hamiltonian is uniform
and unchanging. (This is conventional, for example, in stochastic thermodynamics 
analyses of information-processing systems.)
\item The initial distribution has the property that the vector-valued function $F$ defined by 
\eq{
F(x_A, x_B, x_C) := \left(\ln p_{x_A, x_B}(0) + \ln |X| \;,\; \ln p_{x_C, x_B}(0) + \ln |X| \right)
}
is invertible, as are the associated functions $F_{AB}(x_A, x_B) := \ln p_{x_A, x_B}(0) + \ln |X^{AB}|$ and
$F_{BC}(x_C, x_B) := \ln p_{x_C, x_B}(0) + \ln |X^{BC}|$.
\item $t_f - 0$ is large enough on the scale of the rate matrices so that during the MPP
the joint system relaxes from its (perhaps highly) non-equilibrium initial distribution to the uniform distribution. Note that this does {not} mean that the joint system is in an equilibrium at $t_f$;
depending on the reservoirs coupled to $B$, there might not even be an equilibrium. Rather it means
merely that its distribution then is a fixed point. 
\end{enumerate}

When these conditions are met there is no EF into any of the reservoirs, since the Hamiltonian is uniform. So EP for each
unit $\oo$ is $-\Delta \ln p_{x_\oo}$. In addition, $p_x(t_f)$ is uniform.
Since that distribution is by hypothesis a fixed point
of the forward process, it is also a fixed point of the reverse process. So at the end of the reverse
process, $p(x) := p(x_A, x_B, x_C)$ is uniform. Using the assumed invertibility of $F$ and writing
$|X|$ for the number of states of the joint system, this means that
\begin{widetext}
\eq{
D\left(\vP({\sigma}^{AB}, \sigma^{BC}) \; ||\; {\tilde{\vP}}(-{\sigma}^{AB},- \sigma^{BC})\right)
	&= \sum_{\sigma^{AB}, \sigma^{BC}} \vP(\sigma^{AB}, \sigma^{BC}) 
				\ln \dfrac{\vP(\sigma^{AB}, \sigma^{BC})}{\tilde{\vP}(-\sigma^{AB}, -\sigma^{BC})} \nonumber \\
&= \sum_{\sigma^{AB}, \sigma^{BC}} \left(\sum_{x} p_{x}(0) \delta[F(x), (\sigma^{AB}, \sigma^{BC})]\right)
	\ln \dfrac{\sum_{x} p_{x}(0) \delta[F(x),  (\sigma^{AB}, \sigma^{BC})]}  
			{\sum_{x} |X|^{-1} \delta[F(x), (-\sigma^{AB}, -\sigma^{BC})]}  \nonumber \\
&= \sum_{\sigma^{AB}, \sigma^{BC}} p_{F^{-1}(\sigma^{AB}, \sigma^{BC})}(0) 
							\ln {p_{F^{-1}(\sigma^{AB}, \sigma^{BC})}(0)}   \nonumber \\
& = -S\left(\vP(\sigma^{AB}, \sigma^{BC})\right)
}
\end{widetext}
(Recall the comment below \cref{eq:50_app} about how to interpret reverse process probabilities of negative
EP values.) Similarly,
\eq{
D\left(\vP({\sigma}^{AB}) \; ||\; {\tilde{\vP}}(-{\sigma}^{AB})\right) &= 
		\sum_{\sigma^{AB}} p_{F^{-1}_{AB}}(\sigma^{AB})(0) \ln p_{F^{-1}_{AB}}(\sigma^{AB})(0) \nonumber \\
	& = -S\left(\vP(\sigma^{AB})\right)
}
and
\eq{
D\left(\vP({\sigma}^{BC}) \; ||\; {\tilde{\vP}}(-{\sigma}^{BC})\right) &= 
		\sum_{\sigma^{BC}} p_{F^{-1}_{BC}}(\sigma^{BC})(0) \ln p_{F^{-1}_{BC}}(\sigma^{BC})(0)   \nonumber \\
	& = -S\left(\vP(\sigma^{BC})\right)
}
Combining, we see that when the four conditions given above are met, the multi-divergence 
is just the mutual information between $\sigma^{AB}$ and $\sigma^{BC}$.
%

This establishes the claim in the main text, that the multi-divergence
$I\left(\vP(\sigma^{AB}, \sigma^{BC}) \; ||\; \tilde{\vP}(-\sigma^{AB}, -\sigma^{BC} )\right)$ in the figure caption
is non-negative in many scenarios. The same kind of reasoning can be
extended to establish that many MPPs other than the one considered here also must
have non-negative multi-divergence. (In larger MPPs,
rather than invoke non-negativity of mutual information, as done here, one invokes non-negativity of total correlation.)

To illustrate how the four conditions result in non-negative multi-divergence,
in the rest of this subsection I calculate the multi-divergence explicitly, for a fully specified
initial distribution.
Suppose that systems A and C have the same number of states. Assume further that those
two systems each have a single reservoir, with the same temperature. In addition choose
$p_{x_A, x_B, x_C}(0) = f(x_A) 
\delta(x_A, x_C) \,/\, |X^B|$ for some invertible distribution $f(.)$.

Writing it out, 
\eq{
\sigma^{AB}(\vx) &= -\Delta \ln p_{x_{AB}}(\vx) \nonumber \\
	&=  \ln |X^A| + \ln f(\vx_A(0)) 
\label{eq:e9}
}
and similarly for $\sigma^{BC}(\vx)$. Since $x_A(0) = x_C(0)$ with probability $1$, \cref{eq:e9} means that
$\sigma^{AB} = \sigma^{BC}$ with probability $1$. Since $f$ is invertible, this mean that
\eq{
\vP({\sigma}^{AB}, \sigma^{BC}) &= \delta ({\sigma}^{AB}, \sigma^{BC})e^{ \sigma^{AB} } \;/ \; |X^A|
}
So
\eq{
\vP({\sigma}^{AB}) &= e^{  \sigma^{AB} }  \;/ \; |X^A|
\label{eq:e12}
}

On the other hand, given the dynamics of the reverse process, by \cref{eq:e9},
\eq{
&\tilde{\vP}(-{\sigma}^{AB}, -\sigma^{BC}) = |X^A|^{-2}  \nonumber \\
&\qquad \times \sum_{x_A,x_C} \delta\left(\sigma^{AB}, \ln |X^A| + \ln f(x_A)\right)
					 \delta\left(\sigma^{BC}, \ln |X^A| + \ln f(x_C)\right)
}
(where use was made of the fact that $|X^A| = |X^C|$). So
\eq{
\tilde{\vP}(-{\sigma}^{AB}) &= \sum_{x_A} \delta\left(\sigma^{AB}, \ln |X^A| + \ln f(x_A)\right) \;/\;  |X^A|
}

Plugging these formulas into \cref{eq:e8a} gives
\eq{
&D\left(\vP({\sigma}^{AB}, \sigma^{BC}) \; ||\; {\tilde{\vP}}(-{\sigma}^{AB},- \sigma^{BC})\right)
			   \nonumber \\
&\qquad\qquad\qquad=  -\sum_{\sigma^{AB}}  \dfrac{e^{  \sigma^{AB} }}{|X^A| }\ln \dfrac{|X^A|^{-1}}{ e^{ \sigma^{AB} }}
							\nonumber \\
&\qquad\qquad\qquad=   \sum_{\sigma^{AB}} \dfrac{e^{  \sigma^{AB} }}{|X^A| }\left[\sigma^{AB} +  \ln |X^A| \right]
}
and
\eq{
&D\left(\vP({\sigma}^{AB}) \; ||\; {\tilde{\vP}}(-{\sigma}^{AB})\right) 	 =   
			 \sum_{\sigma^{AB}} \dfrac{e^{  \sigma^{AB} }}{|X^A| }\sigma^{AB} 
}
Combining and using \cref{eq:e12} gives
\eq{
&I\left(\vP({\sigma}^{AB}, \sigma^{BC}) \; ||\; {\tilde{\vP}}(-{\sigma}^{AB}, -\sigma^{BC})\right) 
		= \ln|X^A| \;-\; \langle \sigma^{AB} \rangle 
}
This is always non-negative since the maximal value of $\sigma^{AB}$ under any $f(.)$ is $\ln |X^A|$.

\subsection{Other implications of vector-valued DFT}


\cref{eq:40a} can be elaborated in several way. 
Let $\BBB$ be any subset of the units in $\AAA$. Using the chain-rule for KL divergence to expand
the RHS of \cref{eq:40a}, and then applying \cref{eq:40a} again, this time
with $\AAA$ replaced by $\BBB$, gives
\eq{
\label{eq:41}
\left \langle \sigma^{ \AAA} \right \rangle & =  \left \langle \sigma^{\BBB} \right \rangle +  
	D\left(\vP(\vec{\sigma}^\AAA \,|\, \vec{\sigma}^\BBB) \,||\,  \tilde{\vP}(\vec{\sigma}^\AAA  \,|\, \vec{\sigma}^\BBB) \right) 
}
So the difference in expected total EPs of $\AAA$ and $\BBB$ exactly equals the conditional KL-divergence between the associated EP vectors.

We can also apply the kind of reasoning that led from the vector-valued DFT to \cref{eq:40a}, but after
plugging \cref{eq:global_EP_decomp_in_ex} in the main text  into 
the conditional DFT.
This shows that for any value of $\sigma^\oo$ with nonzero probability,
\eq{
& \widehat{\sum_{\oo'\in {\NN^*}}}  \left\langle  \sigma^{\oo'} \,\bigg|\, \sigma^\oo \right\rangle  - \left\langle \Delta \II^{{\NN^*}}  \,\bigg|\, \sigma^\oo \right\rangle  \nonumber \\
	&\qquad=   \sigma^\oo + D\left(\vP(\vec{\sigma}^{\NN^*} \,\downharpoonright\, \sigma^\oo) 
						\,||\, \tilde{\vP}(-\vec{\sigma}^{\NN^*} \,\downharpoonright\, -\sigma^\oo) \right)
}
(where $\downharpoonright$ indicates conditioning on a specific value of a random
variable rather than averaging over those values, as in the conventional definition of conditional relative entropy).
Similarly, the conditional DFT associated with \cref{eq:32c}
shows that for any value of $\sigma^\oo$ with nonzero probability,
\eq{
\left\langle \chi^\oo| \sigma^\oo \right\rangle &= D\left(\vP(\vec{\sigma}^{\NN^*} \,\downharpoonright\, \sigma^\oo) 
						\,||\, \tilde{\vP}(-\vec{\sigma}^{\NN^*} \,\downharpoonright\, -\sigma^\oo) \right)
}


\section{Sufficient conditions for the conditional mutual information not to increase}
\label{app:conditional_independence_rule}



If $C$ observes $B$ continually, then it may gain some information about the trajectory of $B$
that is not necessarily captured by $B$'s final state.
That information however would tell $C$ something about the ending state
of $A$, if $A$ had also observed the entire trajectory of $B$. 
This phenomenon works to increase $I(A ; C \,|\, B)$ during the process.
On the other hand, in the extreme case where neither $A$ nor $C$ observes
$B$, and the three systems are initially independent of one
another, by the data-processing inequality the mutual information between
$A$ and $C$ will drop during  the process, and therefore so will 
 $\langle \Delta I(A ; C \,|, B)$. So in general, $\langle \Delta I(A ; C \,|, B)$ can be either negative or positive. 
In this appendix I present some simple sufficient conditions for it to be non-positive.

First, note that if both $p_{x_A,x_C \,|\, x_B}(0) \propto \delta(x_A, x_C)$ and
$p_{x_A,x_C}(t_f) = \delta(x_A, x')\delta(x_C,x')$ for some special $x'$,
then $ \Delta I(A; C \,|\, B)  \le 0$. (See~\cref{app:conditional_independence_rule} 
for other cases.) There are also many cases where $ \Delta I(A; C \,|\, B)  > 0$. For example, 
since conditional mutual information is non-negative, that is generically the case if 
$ I(A; C \,|\, B)(0)  =0$ and $x_B$ evolves non-deterministically.

In addition, in App.\,C in~\cite{kolchinsky2020entropy}, it was shown that if any initial distribution in
which $A$ and $C$ are conditionally independent given $B$
is mapped by the MPP to  a final distribution with the same property, then the MPP cannot increase
$I(A ; C \;|\; B)$, no matter the initial distribution actually is. Loosely speaking,
if the MPP ``conserves conditional independence of $A$ and $C$ given $B$'',
then it cannot increase $I(A ; C \;|\; B)$.

The claim was originally proven as part of a complicated
analysis. To derive it more directly, first note that for any joint distribution $p(x_A, x_B, x_C)$,
\eq{
\langle \Delta I_p(A; C \,|\, B) \rangle &= D(p_{X_{ABC}} \;||\; p_{X_{A|B}} p_{X_{C|B}} p_{X_B})
}
where $D(\cdot \;||\; .)$ is relative entropy. rite 
the conditional distribution of the entire MPP as $\Phi(\cdot)$, so 
for any initial distribution $q$, the associated ending distribution is $q(t_f) = \Phi \left(q(0)\right)$.
As shorthand, write $p' = \Phi(p_{X_{ABC}}(0))$.

Since by hypothesis the MPP sends distributions where $x_A$ and $x_C$ are conditionally independent
to distributions with the same property, there must be some distribution $q$ such that
\eq{
\Phi \left(p_{x_{A|B}}(0)p_{x_{C|B}}(0) p_{x_{B}}(0)\right) &= q_{x_{A|B}}(t_f)q_{x_{C|B}}(t_f)q_{x_{B}}(t_f)
}
In addition, since the overall MPP is a discrete time Markov chain, the chain rule for relative
entropy applies. As a result we can expand
\eq{
0 &\leq D\left(p_{A B C} \| p_{A \mid B} p_{C \mid B} p_{B}\right)-D\left(p_{A B C}^{\prime} \| \Phi\left(p_{A \mid B} p_{C \mid B} p_{B}\right)\right) \\
&=D\left(p_{A B C} \| p_{A \mid B} p_{C \mid B} p_{B}\right)-D\left(p_{A B C}^{\prime} \| q_{A \mid B} q_{C \mid B} q_{B}\right) \\
&=D\left(p_{A B C} \| p_{A \mid B} p_{C \mid B} p_{B}\right) - D\left(p_{A B C}^{\prime} \| p_{A \mid B}^{\prime} p_{C \mid B}^{\prime} p_{B}^{\prime}\right) \nonumber \\
&\qquad\qquad -D\left(p_{A \mid B}^{\prime} p_{C \mid B}^{\prime} p_{B}^{\prime} \| q_{A \mid B} q_{C \mid B} q_{B}\right) \\
&\leq D\left(p_{A B C} \| p_{A \mid B} p_{C \mid B} p_{B}\right)-D\left(p_{A B C}^{\prime} \| p_{A \mid B}^{\prime} p_{C \mid B}^{\prime} p_{B}^{\prime}\right)
}
This establishes the claim.

As an illustration, the MPP 
conserves conditional independence of $A$ and $C$ given $B$ (and so by the claim just
established, $\Delta I(A ; C \;|\; B) \le 0$) so long as $x_B$ evolves in an invertible deterministic manner during the MPP.
So in particular, i$\Delta I(A ; C \;|\; B) \le 0$ if $x_B$ does not change its state during the MPP.

%
To see this, as shorthand
write $x_A(t_f) = x'_A, x_A(0) = x_A$, and similarly for systems $B$ and $C$. Write the entire joint
trajectory as $\vx$, as usual. Under the hypothesis that $x_B$ evolves
deterministically, we can write $\vx_B(t) = V_{x_B}(t)$ for some
function $V$. Since that deterministic dynamics of $B$ is invertible, we can also write $V_{x_B}(t) = V'_{x'_B}(t)$
for some function $V'$, and can write $x_B = f(x'_B)$
for some invertible function $f$.

Using this notation, if the initial
distribution $A$ is conditionally independent of $C$ given $B$, then due to the dependency structure we can expand
\begin{widetext}
\eq{
p(x'_A, x'_B, x'_C) &= \sum_{x_A, x_B, x_C} \int \mathcal{D}\vx_A \mathcal{D}\vx_B \mathcal{D}\vx_C \; \delta(\vx_A(t_f), x'_A)
		 \delta(\vx_B(t_f), x'_B)  \delta(\vx_C(t_f), x'_C)  									 \nonumber \\
	&\qquad \qquad\qquad \qquad\qquad p(\vx_A \;|\; \vx_B, x_A) p(\vx_C \;|\; \vx_B, x_C) p(\vx_B  \;|\; x_B)
			p(x_A \;|\; x_B) p(x_C \;|\; x_B) p(x_B)										 \nonumber \\
	&= \sum_{x_A, x_B, x_C} \int \mathcal{D}\vx_A  \mathcal{D}\vx_C \; \delta(\vx_A(t_f), x'_A)
		 \delta(\vx_B(t_f), x'_B)  \delta(\vx_C(t_f), x'_C)   						\nonumber \\
	&\qquad \qquad\qquad \qquad\qquad p(\vx_A \;|\; V_{x_B}, x_A) p(\vx_C \;|\; V_{x_B}, x_C) 
			p(x_A \;|\; x_B) p(x_C \;|\; x_B) p(x_B)										 \nonumber \\
	&= \sum_{x_A, x_C}
p(x'_A \;|\; V'_{x'_B}, x_A) p(x'_C \;|\; V'_{x'_B}, x_C) 
			p(x_A \;|\; f(x_B')) p(x_C \;|\; f(x'_B)) p(f(x'_B))										 \nonumber \\
	&= 
p\left(x'_A \;|\; V'_{x'_B}, f(x'_B) \right) p\left(x'_C \;|\; V'_{x'_B}, f(x'_B) \right) p\left(f(x'_B)\right)	
}
\end{widetext}
which establishes that under the ending distribution, $A$ and $C$ are conditionally independent given $B$, as claimed.


\section{Cases where system LDB is exact under global LDB}
\label{app:exact_ldb}

In this paper I assume that \cref{eq:nldb} holds, and that 
all fluctuations in the state of system $i$ are due to exchanges with its reservoir(s), and that the amounts
of energy in such exchanges are given by the associated changes in the value of $i$'s  local 
Hamiltonian. (This is the precise definition of SLDB). In this appendix I illustrate a generic
scenario in which these assumptions all hold.

\subsection{Hamiltonian stubs}

Rather than start with a set of rate matrices, one per system, start with an additive decomposition of
the global Hamiltonian into a set of $D$ different terms, where each term $d \in \{1, \ldots, D\}$ only depends on the joint
state of the systems in some associated subset of $\NN$. Specifically, for each $d \in \{1, \ldots, D\}$,
choose some arbitrary set ${m(d)} \subseteq \NN$, along with an associated arbitrary \textbf{Hamiltonian stub}, $h_{d, t}(x_{m(d)})$,
and define the global Hamiltonian to be a sum of the stubs:
\eq{
H_x(t) = \sum_{d=1}^D h_{d, t}(x_{m(d)})
\label{eq:global_hamiltonian}
}
Next, for all systems $i$, define the associated neighborhood and local Hamiltonian by
\eq{
r(i) &:= \cup_{d : i \in m(d)} m(d) \\
H_{x}(i; t) &:= H_{x_{r(i)}}(i; t) \\
	&:= \sum_{d : i \in m(d)} h_{d, t}(x_{m(d)})
\label{eq:local_hamiltonian}
}

Note that for an MPP, global, exact LDB says that fluctuations in the state of system $i$ due
to its reservoirs are governed by a Boltzmann distribution with Hamiltonian $H_{\vx(t)}(i; t)$.
Note also that the global Hamiltonian does not equal the sum of the local Hamiltonians in general, i.e., 
it may be that
\eq{
\sum_{d=1}^D h_{d, t}(x_{m(d)}) \ne \sum_i \sum_{d : i \in m(d)} h_{d, t}(x_{m(d)})
}
On the other hand, since in an MPP only one system changes state at a time,
the total heat flow to the reservoirs of all the systems along any particular trajectory $\vx$
\textit{is} the sum of the associated changes in the values of the local Hamiltonians (which are given in \cref{eq:local_hamiltonian}). 
For the same reason, SLDB holds \textit{exactly}, without any approximation error.

\subsection{Example involving diffusion between organelles}


Consider a pair of systems, $A, C$, with an intervening ``wire'', $b$,
that diffusively transports signals from $A$ to $C$, but not vice-versa. It is meant to be an abstraction
of what happens when one cell sends a signal to another, or one organelle within a cell sends a 
signal to another organelle, etc. So $C$ is observing $A$, but
$A$'s dynamics is independent of the state of $C$. 

To formalize this scenario, let $Y$ be some finite space with at least three elements, with a special element labeled $0$.
Let $X_A$ be a vector space, with components $\{x_{A}(i) : i = 1, \ldots\}$, where
$X_{A}(1) = Y$. We will interpret $x_A(1)$ as the ``emitting signal''. 
Let $X_b = Y^m$ for some $m \ge 3$, and write those $m$ components of $x_b$
as $x_b(1), x_b(2), \ldots x_b(m)$. Intuitively, the successive components of $b$ are successive positions on a physical wire, stretching
from $A$ to $C$; if $x_{b}(j) = 0$, then there is no signal at location $j$ on the wire. 
Finally, let $X_C$ also be relatively high-dimensional, with components $\{x_{C}(i) : i = 1, \ldots\}$,
where $X_{C}(1) = Y$. We will interpret $x_C(1)$ as the ``received signal''. 

The key to getting this setup to model a signal propagating from $A$ to $C$ lies
in the Hamiltonian. Have three Hamiltonian stubs
which are combined as in \cref{eq:global_hamiltonian} to define that global Hamiltonian.
For clarity, indicate those three stubs with the letters $A, b$ and $C$, respectively, rather than the three numbers $d \in \{1, 2, 3\}$.
Also make the definitions in \cref{eq:local_hamiltonian}. So global LDB automatically
ensures SLDB. 
From now on, to reduce notation clutter, the time index $t$ will be implicit.

Choose $m(A) = \{A\}, m(b) = \{A, b, C\}$, and $m(C) = \{b, C\}$. Therefore $r(A) = \{A, b\}, r(b) = \{A, b, C\}$, and $r(C) = \{b, C\}$.
Also choose $h_C$ to only depend on the $m$'th component of $x_b$, and
choose $h_b$ to only depend on the first components of $x_A$ and $x_C$. So
we can write the stub of $C$ as $ h_C(x_b(m), x_C)$ and write the stub of $B$ as
$h_b(x_A(1), x_b, x_{C}(1))$.

Combining, the local Hamiltonians of the three systems are
\eq{
\label{eq:full_A_Hamiltonian}
H_{x_A, x_b, x_C(1)}(A) &= h_A(x_A) + h_b(x_A(1), x_b, x_{C}(1)) \\
H_{x_A(1), x_b, x_C(1)}(b) &= h_A(x_A) + h_b(x_A(1), x_b, x_{C}(1)) \nonumber \\
	&\qquad\qquad + h_C(x_b(m), x_C) \\
H_{x_{A}(1), x_b(m), x_C(1)}(C) &= h_b(x_A(1), x_b, x_{C}(1)) + h_C(x_b(m), x_C) 
\label{eq:full_C_Hamiltonian}
}
Finally, assume we can write
\eq{
h_C(x_b(m), x_C) = \tilde{h}^C(x_C) + \tilde{h}^{bC}(x_b(m), x_{C}(1))
\label{eq:e34}
}
for some functions $\tilde{h}^C, \tilde{h}^{bC}$ such that for any value of $x_b(m)$ and $x_{C}(1) \ne x_b(m)$,
\eq{
\tilde{h}^{bC}(x_b(m), x_b(m)) < \tilde{h}^{bC}(x_b(m), x_C(1))
\label{eq:e35}
}

For simplicity assume that at most one signal can be on the wire at a given time. So  $h_b(x_A(1), x_b, x_{C}(1)) = \infty$
if more than one component of $x_b$ differs from $0$. $h_b(x_A(1), x_b, x_{C}(1))$ has the same fixed value,
$E$, for all other values of its arguments. This means that $h_b(x_A(1), x_b, x_{C}(1))$ is independent
of the values of $x_A(1)$ and $x_C(1)$. So as far as all rate matrices are concerned, by LDB
we can rewrite \cref{eq:full_A_Hamiltonian,eq:full_C_Hamiltonian} as
\eq{
\label{eq:e36}
H_{x_A}(A) &= h_A(x_A) \\
H_{x_b(m), x_C(1)}(C) &= h_C(x_b(m), x_C) 
\label{eq:e37}
}
\cref{eq:e36} establishes that the dynamics of $A$ is autonomous, independent of the 
states of $B$ and $C$, as claimed. Given \cref{eq:e34,eq:e35}, \cref{eq:e37} 
establishes that the only dependence of the dynamics of $C$ on the state of $b$ is
a bias in favor of overwriting $x_C(1)$ with $x_b(m)$. 

%
We impose extra restrictions on $K(b)$ in addition to LDB:
\begin{enumerate}
\item 
\label{item:constraint1}
Suppose both $x'_b$ and $x_b$ have exactly one nonzero component, with indices $j', j \ne j'$,
respectively, and write
$x'_b(j') = y'$ and $x_b(j) = y$. (So the nonzero component of $x'_b$ is not its last component.) Then 
\eq{
K^{x_A(1), x'_b, x_{C}(1)}_{x_A(1), x_b, x_{C}(1)}(b) = 0
} 
if $y' \ne y$. (So the signal in the wire cannot spontaneously change.) If $y = y'$ however, and $|j-j'| = 1$,
then 
\eq{
K^{x_A(1), x'_b, x_{C}(1)}_{x_A(1), x_b, x_{C}(1)}(b; t) \ne 0
}
(So the signal can diffuse across the wire.)
\item 
\label{item:constraint2}
If $x'_b = \vec{0}$ (i.e., all components of $x'_b$ equal $0$), and $x_b(1) = x_A(1) \ne 0$ then
\eq{
K^{x_A(1), x'_b, x_{C}(1)}_{x_A(1), x_b, x_{C}(1)}(b) \ne 0
\label{eq:e35aa}
}
(So the wire can copy the signal in $x_A(1)$ into its first position.)
However, if instead $x'_b = \vec{0}$ and $x_A(1) \ne 0$ but $x_b(1) \ne x_A(1)$, then 
\eq{
K^{x_A(1), x'_b, x_{C}(1)}_{x_A(1), x_b, x_{C}(1)}(b)= 0
}
(So the wire cannot make a mistake when copying the signal in $x_A(1)$ into its first position.)
\item 
\label{item:constraint3}
If $x'_b(m) = x_C(1)$ and $x_b = \vec{0}$, then
\eq{
K^{x_A(1), x'_b, x_{C}(1)}_{x_A(1), x_b, x_{C}(1)}(b) \ne 0
}
So if the signal in $x_b(m)$ has already been copied into $x_C(1)$ --- or serendipitously already exists
in $x_C(1)$ --- then $x_b(m)$ can be reset to $0$. Such a resetting will allow a new signal to enter
the wire from $A$.
\end{enumerate}
All other terms in the matrix $K(b)$ that are not set by either LDB or normalization equal $0$. 

Finally, given \cref{eq:e35aa}, we can also choose to $h_C(x_b(m), x_C)$ controls the dynamics of $C$, with a bias in favor
of copying the value of $x_b(m)$ into $x_C(1)$. 

Note that the set of rate matrix constraints on $K(b)$ is consistent
with LDB, since all states of the wire with nonzero probability have the same energy value, $E$.
In particular, LDB can be consistent with \cref{eq:e35aa}; it simply means that if the wire has a single nonzero entry,
in its first location, which happens to equal $x_A(1)$, then the wire can ``fluctuate'' into the state $\vec{0}$,
losing all information.

\end{document}